\documentclass[useAMS,usenatbib,onecolumn]{mn2e}

\input tikz
\usetikzlibrary{decorations.markings}

\newcommand{\kp}{\mbox{\tiny{K}}}
\newcommand{\p}{\mbox{\scriptsize{p}}}
\newcommand{\A}{\mbox{\tiny{A}}}
\newcommand{\mm}{\mbox{\scriptsize{m}}}
\newcommand{\co}{\mbox{\scriptsize{co}}}
\newcommand{\s}{\mbox{\scriptsize{s}}}

\newcommand{\ac}{\mbox{\scriptsize{a}}}
\newcommand{\ssl}{\mbox{\scriptsize{sl}}}
\newcommand{\crit}{\mbox{\scriptsize{crit}}}

\newcommand{\sn}{\mbox{\scriptsize{sn}}}
\newcommand{\eq}{\mbox{\scriptsize{eq}}}

\usepackage{amssymb,amsmath}
\begin{document}

\title{Angular momentum transport through magnetic accretion curtains inside disrupted discs}

\author[C. G. Campbell]{C. G. Campbell \\ School of Mathematics and Statistics, Herschel Building, Newcastle 
University, NE1 7RU}

\date{Accepted for publication in MNRAS}

\maketitle

\begin{abstract}
 The structure of accretion curtain flows, which form due to disc disruption by a strongly magnetic star, is considered.
It is shown that a sub-Alfv\'enic, magnetically channelled flow is consistent with matching the magnetic field across
the curtain base where it meets the disrupted inner region of the disc. The resulting angular velocity distribution in
the curtain flow is calculated, together with the consequent angular momentum tranfer rate to the star.

 It is shown that the transition of material from the diffusive disc flow to the channelled curtain flow results in some
angular momentum being fed back into the disc. This is consistent with the total angular momentum balance, and can result
in a significantly smaller accretion torque acting on the star than that given by the standard model which assumes no
angular momentum feed-back to the disc. The sonic point coordinates are found and a critical stellar rotation rate results below which the sonic point merges with the curtain flow base, due to a reduced centrifugal barrier to the flow. Hence, at 
these lower rotation rates, little thermal assistance is required for matter to make the transition to a supersonic accretion flow.
\end{abstract}

 
\section{Introduction}

\subsection{Types of systems}

 An accretion curtain forms when a strongly magnetic star disrupts the inner part of its surrounding accretion disc. 
Material leaves the disc and becomes channelled by the stellar magnetic field until it reaches the star's surface, via localised columns of shock-compressed gas. The angular momentum transferred in the curtain flow plays an important 
role in determining the spin evolution of the star. Such accretion is believed to occur onto a protostar in T Tauri systems (e.g. Bouvier et al 2007), a neutron star in X-ray binary pulsars (e.g. Campbell 1997) and a white dwarf in the
intermediate polars (e.g. Warner 1995; Frank, King \& Raine 2002).

 The intensity of radiations emitted by the shocked hot gas in the accretion columns is rotationally modulated, and
observations of this yield information on the spin behaviour of the star (e.g. Chakrabarty 1997). Spin-up and spin-down
are observed, as well as more erratic period changes (e.g. Bildsten et al 1997). The nature of the curtain flow and the
associated angular momentum transfer are of fundamental significance in these important systems.

\subsection{Previous Work} 

 Pringle \& Rees (1972) suggested that a strongly magnetic neutron star accreting via a disc is likely to disrupt
the disc far from the stellar surface inside a corotation radius, $\varpi_{\co}$, at which the Keplerian angular velocity 
of material in the central plane of the disc equals the stellar angular velocity. They argued that inside the disruption
radius ionized matter would be channelled by the strong magnetic field of the star before being accreted on to its
surface.

 Ghosh \& Lamb (1978, 1979a,b) constructed a basic model to describe the interaction of a magnetic neutron star with
its accretion disc. They pointed out that the stellar magnetic field is likely to penetrate the thin turbulent disc
as a consequence of instabilities and reconnection. The magnetosphere is expected to nearly corotate with the star,
and the vertical shear between this surrounding region and the disc material leads to the generation of toroidal
magnetic field from the poloidal component. The resulting magnetic stresses lead to angular momentum exchange between
the disc and star. The toroidal magnetic field reverses sign across $\varpi_{\co}$, and hence the direction of angular momentum exchange between the disc and star also reverses. Outside $\varpi_{\co}$ the star adds angular momentum to 
the disc, while inside this radius angular momentum is removed from the disc. Hence these regions make spin-down and 
spin-up contributions, respectively, to the stellar torque.

 It was sugessted by Ghosh \& Lamb that magnetic stresses would result in disc disruption in the region of a spherical 
Alfv\'en radius, $\varpi_{\A}$, inside $\varpi_{\co}$. It was argued that disruption should occur over a narrow annulus,
through which material leaves the disc vertically and enters a channelled flow guided by the relatively
strong stellar magnetic field. However, no mechanism was given to explain disruption of the disc or the vertical
transition of the disc flow into the field-controlled accretion region. The angular velocity of disc material was
taken to change from a Keplerian value to the stellar value across the inner boundary layer in which disruption
occurs. All the angular momentum of material leaving this layer was assumed to be transferred to the star via the
channelled flow and a consequent accretion torque was calculated. A stellar spin rate can be found at which the magnetic
and accretion torques balance, so it is possible to have zero total torque acting on the star while it is accreting
material from the disrupted disc. 

 Campbell (1992) considered the inner part of the disc, inside $\varpi_{\co}$. It was shown that the vertical equilibrium
cannot be maintained if the magnetic torque significantly exceeds the viscous torque. This work indicated that disc
disruption is due to a thermal process which causes the disc to expand vertically over a narrow radial region.

 Most authors have assumed, like Ghosh \& Lamb, that all the angular momentum of the magnetically channelled 
material is transferred to the star (e.g. Wang 1987; Armitage \& Clarke 1996; Matt \& Pudritz 2005; Tessema 
\& Torkelsson 2010; Campbell 2011). However, Shu et al (1994a,b) and Ostriker \& Shu (1995) developed a modified model 
of the star-disc interaction in which the accretion torque on the star is greatly reduced. In this model the field
lines thread the disc, but disruption occurs close to the corotation radius. As in the Ghosh \& Lamb model, matter is
channelled onto the star through a magnetic field guided flow orginating from a thin disruption region in the disc.
However, a wind flow is taken to develop from the disc just beyond this region and this carries a significant amount of
angular momentum away from the system. Consequently the accretion torque delivered to the star is significantly reduced.
The accretion and wind flows emanate from an 'X-type' region centred on $\varpi_{\co}$, in which it is assumed that a
steady state can be maintained. This model was constructed for particular application to protostellar systems (T Tauri
stars).

 Li, Wickramasinghe \& Ruediger (1996) argued that the accretion torque could be effectively zero due to all the
angular momentum of the field channelled flow being fed back into the disc. They argued that the transfer of angular
momentum via the channelled flow would require large values of the field winding ratio $|B_\phi/B_{\p}|$ at the
stellar surface, which would lead to instabilities. However, Wang (1997) pointed out that only small values of this
ratio are required at the star's surface since the torque is related to the stress $B_\phi B_{\p}$ which is sufficiently
large even though $|B_\phi/B_{\p}|$ is small.

 Numerical simulations of the field channelled curtain flow have been performed, with one of the most extensive by 
Romanova et al (2002). These simulations showed that funnel flows develop in the inner disc region. Material becomes 
channelled by the stellar magnetic field and this leads to a spin-up torque on the star. The case of a dipolar field was 
considered, with the magnetic moment aligned with the spin axis of the star. Cases with tilted dipoles and multipole 
components were analysed numerically by Romanova et al (2003). These calculations were done in the context of T Tauri 
stars, in which observations suggest that the magnetic field is more complex than dipolar (e.g. Jardine et al 2002).

 Erkut \& Alpar (2004) derived a form for the dynamical viscosity from the steady angular momentum balance for a
Keplerian disc. They then used this in the angular momentum equation to estimate deviations of $\Omega$ from $\Omega_{\kp}$
inside $\varpi_{\co}$. For sufficiently slow values of the stellar rotation rate, $\Omega_*$, they found that $\Omega$
can change from $\Omega_{\kp}$ to $\Omega_*$ over a broader region than the boundary layer suggested by Ghosh \& Lamb.
However, for larger values of $\Omega_*$ boundary layer behaviour is exhibited. A detailed disc solution was not found,
nor a mechanism for disc disruption. 

 It was shown in Campbell (2010) that disc disruption occurs inside the corotation radius as a result of increasing
magnetic heating.  The consequent growing central plane pressure cannot be balanced by the vertical component of the 
stellar gravity and the disc expands over a narrow radial region. At disruption the magnetic torque is rapidly increasing and material is diverted from the disc into a magnetically channelled region, forming a curtain-type flow. The increased
vertical pressure gradients allow material to surmount the effective potential barrier and flow through the sonic point. 
Beyond this point, the component of the stellar gravity along the poloidal magnetic field exceeds that due to the centrifugal force and material is subsequently accelerated to speeds approaching free-fall values. An approximate expression was found for the position of the sonic point, and the energy required to reach it. 

\subsection{Paper outline}

 The present paper looks at the nature of the accretion flow in more detail than previous studies. In Section 2 the
magnetic and dynamical problems are addressed. The equations are solved for the flow and magnetic field, as a function
of the stellar rotation rate. Matching of the disc and curtain structures is shown to have fundamental consequences for
the angular momentum transport to the star. 

 In Section 3 the slow magnetosonic point is considered and its coordinates are calculated. This is shown to be essentially
the same as the sonic point. A critical stellar rotation rate results at which the slow point approaches the disc surface,
hence facilitating the transition to supersonic flow. The resulting accretion torque is calculated in Section 4, as a
function of the stellar rotation rate. This torque can be significantly lower than that given by the standard form, since
some angular momentum can be transferred back to the disc in the transition layer at its inner edge. Section 5 discusses 
the results.

\section{The Governing Equations}

 A dipolar magnetic field is considered since, in the region of interest, this will be the dominant component in white
dwarf and neutron star systems. In general, the magnetic moment will be inclined to the rotation axis but, as illustrated
in Campbell (2011), the magnetic field can be split into time-dependent and time-independent parts with the latter being
axisymmetric. The time-independent field dominates in the relevant part of the accretion flow and, in cylindrical
coordinates $(\varpi,\phi,z)$, its unperturbed components are
$$
B_\varpi=\frac{3}{2}B_0R^3\frac{\varpi z}{(\varpi^2+z^2)^{5/2}}, \qquad
B_z=-\frac{1}{2}B_0R^3\frac{(\varpi^2-2z^2)}{(\varpi^2+z^2)^{5/2}},
\eqno{(1\mbox{a,b})}
$$
where $B_0$ and $R$ are the polar magnetic field strength and the stellar radius, respectively. It will be shown that
the accretion flow only causes small distortions of the stellar magnetic field from this form.

 The steady magnetohydrodynamic equations will be employed for the curtain flow, in the limit of high magnetic Reynolds
number. The quantities conserved along poloidal magnetic field lines, that are presented in this section, are similar to 
those occurrring in a steady, axisymmetric magnetic wind flow (e.g. Campbell 1997, Mestel 1999 and references therein). 
However, some important differences arise since the poloidal velocity in the accretion curtain is a sub-Alfv\'enic inflow,
as opposed to the trans-Alfv\'enic outflow associated with a magnetically influenced wind. The poloidal and toroidal components of the induction equation yield
$$
{\bf v}_{\p}=\kappa{\bf B}_{\p},
\eqno{(2)}
$$
$$
\Omega-\frac{\kappa B_\phi}{\varpi}=\alpha,
\eqno{(3)}
$$
where $\kappa$ is a function of position, $\Omega$ is the angular velocity of material and $\alpha$ is the angular velocity
of a poloidal field line. Since the poloidal field originates in the degenerate accretor, where it is frozen to the highly
conducting stellar material, the constant $\alpha$ is also the stellar angular velocity $\Omega_*$. Conservation of mass
and magnetic flux lead to the quantity 
$$
\epsilon=\rho\kappa=\frac{\rho v_{\p}}{B_{\p}},
\eqno{(4)}
$$
being conserved along poloidal field-streamlines. It is noted that, since ${\bf v}_{\p}$ and ${\bf B}_{\p}$ are 
anti-parallel in the accretion curtain, for $z>0$, $\kappa$ and $\epsilon$ are negative.

 Another quantity conserved along field-streamlines follows from the angular momentum equation as 
$$
\epsilon\varpi^2\Omega-\frac{\varpi B_\phi}{\mu_0}=-\beta.
\eqno{(5)}
$$
This gives the total rate of transport of angular momentum, carried jointly by the gas and magnetic stresses. Equations
(3)--(5) combine to yield
$$
\Omega=\frac{\Omega_*-\mu_0|\epsilon|\beta/\varpi^2\rho}{1-\mu_0\epsilon^2/\rho},
\eqno{(6)}
$$
and
$$
B_\phi=\frac{\mu_0\beta/\varpi-\mu_0|\epsilon|\Omega_*\varpi}{1-\mu_0\epsilon^2/\rho}.
\eqno{(7)}
$$
It follows from (4) that
$$
\frac{\mu_0\epsilon^2}{\rho}=\frac{\rho v^2_{\p}}{B^2_{\p}/\mu_0}=\frac{v^2_{\p}}{v^2_{\A}},
\eqno{(8)}
$$
where $v_{\A}$ is the Alfv\'en speed. The dimensionless quantity $v^2_{\p}/v^2_{\A}$ gives the ratio of the poloidal
kinetic energy density to the poloidal magnetic energy density. This will be shown to be of fundamental importance
to the properties of the curtain flow.

 The flow is taken to be isothermal, with sound speed $a$. Consideration of the rate of work done by the forces acting
on the gas leads to the generalized Bernoulli integral
$$
\frac{1}{2}v^2_{\p}+\frac{1}{2}\varpi^2\Omega^2-\frac{GM}{(\varpi^2+z^2)^{1/2}}+a^2\ln\rho-\Omega_*\varpi^2\Omega=E.
\eqno{(9)}
$$
This represents the sum of the poloidal and toroidal kinetic energies, the gravitational energy, and the work done by the
pressure gradient and the magnetic torque, all per unit mass. The total energy $E$ is conserved along poloidal magnetic 
field lines on which $z=z(\varpi)$.

\section{Conditions at the curtain flow base}

\subsection{The poloidal flow speed}

 The disc and curtain flows must be matched in the region of the disruption radius. It was shown in Campbell (2010)
that the disc becomes disrupted over a narrow region, of width $\delta$, inside a radius of $\varpi_{\mm}$ given by
$$
\varpi_{\mm}=\left(\frac{2\gamma}{\epsilon_{\mm}}\right)^{2/7}(1-\xi^{3/2})^{2/7}\varpi_{\A},
\eqno{(10)}
$$
where $\gamma$ and $\epsilon_{\mm}$ are dimensionless quantities related to the disc vertical shear distribution and 
diffusivity, while $\varpi_{\A}$ is a spherical Alfv\'en radius, with
$$
\varpi_{\A}=\left[\frac{\pi B^2_0R^6}{\mu_0(GM)^{1/2}{\dot M}}\right]^{2/7}.
\eqno{(11)}
$$
Values of $\gamma/\epsilon_{\mm}\sim 1$ ensure $|B_\phi/B_z|<1$ at the base of the curtain flow, this being consistent
with strong field channelling in a sub-Alfv\'enic flow. The stellar rotation parameter $\xi$ is
$$
\xi=\frac{\varpi_{\mm}}{\varpi_{\co}}=\left(\frac{\Omega_*}{\Omega_{\kp\mm}}\right)^{2/3},
\eqno{(12)}
$$
where the corotation radius is
$$
\varpi_{\co}=\left(\frac{GMP^2}{4\pi^2}\right)^{1/3},
\eqno{(13)}
$$
with $P=2\pi/\Omega_*$, $\Omega_{\kp\mm}=\Omega_{\kp}(\varpi_{\mm})$ and 
$$
\Omega_{\kp}=\left(\frac{GM}{\varpi^3}\right)^{1/2}.
\eqno{(14)}
$$
The disc disruption conditions lead to
$$
\xi=\left[\frac{1}{2}\left\{(C^2+4C)^{1/2}-C\right\}\right]^{2/3},
\eqno{(15)}
$$
with
$$
C=2\frac{\gamma}{\epsilon_{\mm}}\left(\frac{\varpi_{\A}}{\varpi_{\co}}\right)^{7/2}.
\eqno{(16)}
$$
Disruption occurs inside $\varpi_{\co}$ (i.e. $\xi<1$) and is due to the rapidly growing central plane pressure caused by
magnetic heating increasing with decreasing distance from the star. The disc expands vertically and the accretion 
flow is fed into the magnetically controlled curtain through its base, which is a ring of width $\delta$ situated
just inside $\varpi_{\mm}$, with $\delta \ll \varpi_{\mm}$. Hence this thin transition region has a radial extent of 
$\varpi_{\mm}-\delta<\varpi<\varpi_{\mm}$ and a field line through its centre, with equation $z=z(\varpi)$, is taken to 
have the base coordinates $\varpi=\varpi_0$, $z=z(\varpi_0)=h_0$. To first order in $h_0/\varpi_0$, this field line is 
vertical in the disc and hence the coordinate $\varpi_0$ at which it cuts the disc surface can be taken to be the same as 
the coordinate $\varpi_{\mm}$ at which it cuts the disc mid-plane $z=0$, for most purposes. 
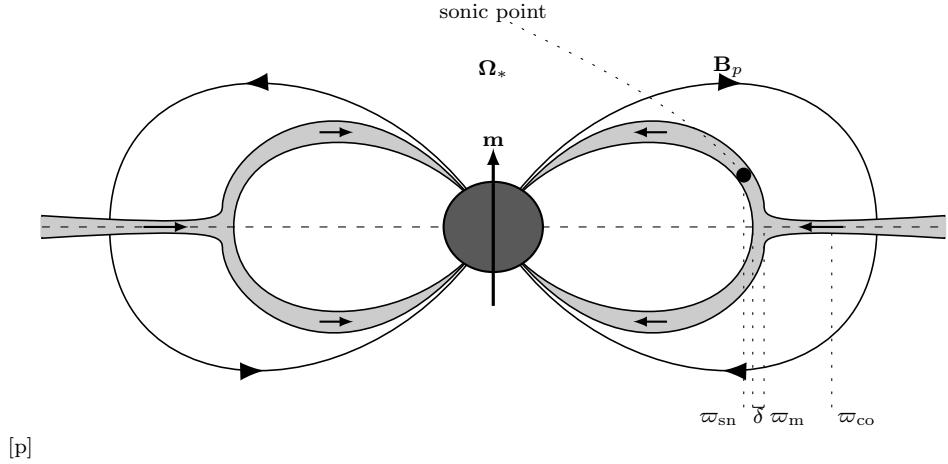
\begin{figure}[p]
\centering
\begin{tikzpicture}[domain=-2:2, scale=3, samples=100, x=1cm]

	  \draw[white]
	  	( 0.00,-1.00) -- ( 1.20,-0.80) node(xline) {};

	  \draw[semithick,decoration={markings,mark=at position 0.58 with {\arrow[scale=2]{latex}}}, postaction={decorate}]
		( 0.00, 0.00) .. controls ( 0.51, 0.85) and ( 1.70, 0.85) .. 
			node[pos=0.5,above=2pt] {$\mathbf{B}_p$} ( 1.70, 0.00);
	  \draw[semithick,decoration={markings,mark=at position 0.46 with {\arrow[scale=2]{latex}}}, postaction={decorate}]
		( 1.70, 0.00) .. controls ( 1.70,-0.85) and ( 0.51,-0.85) .. ( 0.00, 0.00);
	  \draw[semithick,decoration={markings,mark=at position 0.58 with {\arrow[scale=2]{latex}}}, postaction={decorate}]
		( 0.00, 0.00) .. controls (-0.51, 0.85) and (-1.70, 0.85) .. (-1.70, 0.00);
	  \draw[semithick,decoration={markings,mark=at position 0.46 with {\arrow[scale=2]{latex}}}, postaction={decorate}]
		(-1.70, 0.00) .. controls (-1.70,-0.85) and (-0.51,-0.85) .. ( 0.00, 0.00);
	  \filldraw[fill=gray!40,semithick]%
	  	( 0.00, 0.00)	.. controls ( 0.55,-0.80) and ( 1.17,-0.40) .. 
		( 1.20,-0.10) .. controls ( 1.20,-0.04) and ( 1.20, 0.00) ..
		( 2.00,-0.05) --
		( 2.00, 0.05)	.. controls ( 1.20, 0.00) and ( 1.20, 0.04) ..
		( 1.20, 0.10)	.. controls ( 1.17, 0.40) and ( 0.55, 0.80) ..
		( 0.00, 0.00)	.. controls (-0.55,-0.80) and (-1.17,-0.40) .. 
		(-1.20,-0.10) .. controls (-1.20,-0.04) and (-1.20, 0.00) ..
		(-2.00,-0.05) --
		(-2.00, 0.05)	.. controls (-1.20, 0.00) and (-1.20, 0.04) ..
		(-1.20, 0.10)	.. controls (-1.17, 0.40) and (-0.55, 0.80) ..
		(-0.00, 0.00);
	  \draw[very thick,gray!40]
	  	(-2.00, 0.045) -- (-2.00,-0.045)
		( 2.00, 0.045) -- ( 2.00,-0.045);
	  \filldraw[semithick, fill=white]			
		( 0.00, 0.00) .. controls ( 0.30, 0.50) and ( 1.15, 0.50) ..
		( 1.15, 0.00) .. controls ( 1.15,-0.50) and ( 0.30,-0.50) ..
		( 0.00, 0.00) .. controls (-0.30, 0.50) and (-1.15, 0.50) ..
		(-1.15, 0.00) .. controls (-1.15,-0.50) and (-0.30,-0.50) ..
		( 0.00, 0.00) -- cycle;
		
	  \draw[thin, dash pattern = on 3pt off 4pt]
	  	(-2.00, 0.00) -- ( 2.00, 0.00);	
	  \draw[thick, -latex] (-1.55, 0.00) -- (-1.35, 0.00);
	  \draw[thick, -latex] ( 1.55, 0.00) -- ( 1.35, 0.00);
	  \draw[thick, -latex]
	  	( 0.77, 0.42) -- ( 0.62, 0.42);
	  \draw[thick, -latex]
	  	(-0.77, 0.42) -- (-0.62, 0.42);
	  \draw[thick, -latex]
	  	( 0.77,-0.42) -- ( 0.62,-0.42);
	  \draw[thick, -latex]
	  	(-0.77,-0.42) -- (-0.62,-0.42);

	  \filldraw[draw=black,thick,fill=gray!70!black]
	  	( 0.00, 0.00) ellipse (0.22 and 0.20);
	  \draw[very thick, -latex]
	  	( 0.00,-0.35) -- ( 0.00, 0.35) node[above] {$\mathbf m$};

	  \node[circle,fill=black,inner sep=2pt]
	  	(s)	at ( 1.11, 0.23) {};
	  \node
		(u)	at ( 1.20, 0.00) {};
	  \node
	  	(co)	at ( 1.50, 0.00) {};
	  \node
	  	(t)	at ( 1.15, 0.00) {};
	  \draw[dash pattern=on 1pt off 4pt]
	  	(s |- xline) -- node[below left=2pt, pos=0.0] {$\varpi_{\sn}$} (s)
		(t |- xline) -- (t)
		(u |- xline) -- node[below right=2pt, pos=0.0] {$\varpi_{\mm}$} (u)
		(co |- xline) -- node[below right=2pt, pos=0.0] {$\varpi_{\co}$} (co);
	  \draw
	  	(t |- xline) -- (u |- xline) node[below,pos=0.5] {$\delta$};
		
	  \node
	  	(t1)	at ( 0.00, 0.95) {\small sonic point};
	  \node
	  	(om)	at ( 0.00, 0.70) {$\boldsymbol{\Omega}_*$};
	  \draw[dash pattern=on 1pt off 4pt]
	  	(t1.330) .. controls +(330:0.1) and +(130:0.4) .. (s.130);

\end{tikzpicture}
\caption{Schematic picture of a typical system. The vertical cross-section shows the relative positions of the disc 
corotation radius, $\varpi_{\co}$, the disruption radius, $\varpi_{\mm}$, and the sonic point radius, $\varpi_{\sn}$, in the curtain flow. Disruption occurs over a narrow annulus of width $\delta$ through which matter flows from the disc into the magnetically channelled accretion curtain.}
\end{figure}

 As an illustration, consider $\gamma/\epsilon_{\mm}=1$ and typical neutron star X-ray binary parameters $M=1.4\,M_\odot$,
$R=10^4\,{\mbox m}$, ${\dot M}=2\times 10^{-9}\,M_\odot\,{\mbox{yr}}^{-1}$, $B_0=2\times 10^8\,{\mbox T}$, and
$P=3\,{\mbox s}$. Equations (10)--(16) then lead to $\xi=0.8$, $\varpi_{\mm}=0.8\,\varpi_{\co}$, $\varpi_{\A}
=0.9\,\varpi_{\co}$, with $\varpi_{\co}=3.5\times 10^6\,{\mbox m}$. Fig. 1 shows a general schematic illustration of
such a system.
 
 At disruption, the flow divides evenly between the upper and lower halves of the disc, so only the upper half need be considered. Matter is fed in to the upper curtain through its base at a rate 
$$
\frac{\dot{M}}{2}=2\pi\varpi_{\mm}(\rho v_z)_0\delta,
\eqno{(17)}
$$
with the poloidal field being nearly vertical here, so $v_{\p}=v_z$. The ratio $(v^2_{\p}/v^2_{\A})_0$ can be written as
$$
\left(\frac{v^2_{\p}}{v^2_{\A}}\right)_0=\left(\frac{\rho v^2_z}{B^2_z/\mu_0}\right)_0
=\frac{\varpi_{\mm}}{\delta}\left(\frac{\varpi_{\mm}}{\varpi_{\A}}\right)^{7/2}
\left(\frac{v_z}{v_{\kp}}\right)_0,
\eqno{(18)}
$$
where the last expression follows from using (1b) for $B_z$ and  (17) to eliminate $(\rho v_z)_0$, then employing (11)
for $\varpi_{\A}$ to eliminate ${\dot M}$. This expression can be written as
$$
\left(\frac{v^2_{\p}}{v^2_{\A}}\right)_0=\left(\frac{\varpi_{\mm}}{\varpi_{\A}}\right)^{7/2}
\left(\frac{\varpi_0}{h_0}\frac{a}{v_{\kp 0}}\right)\frac{h_0}{\delta}\frac{v_{z0}}{a}.
\eqno{(19)}
$$
It was shown in Campbell (2010) that $\varpi_{\mm}/\varpi_{\A}\lesssim 1$ and $h_0 \lesssim \delta \ll \varpi_{\mm}$.
As the disc expands vertically in the disruption region hydrostatic equilibrium starts to break down and matter enters the accretion curtain with a subsonic speed $v_{z0}<a$. The nearly vertical force balance in the disc, together with 
$a<c_{\s}$ due to the surface temperature being less than the central plane temperature, gives
$$
\frac{\varpi_0}{h_0}\frac{a}{v_{\kp 0}}\lesssim 1.
\eqno{(20)}
$$
It therefore follows from (19) that the initial curtain flow is well sub-Alfv\'enic.

 Evaluating the conserved quantity $\rho v_{\p}/B_{\p}$ at the disc surface and the sonic point leads to
$$
\frac{v_{z0}}{a}=\frac{|B_z(\varpi_0)|}{B_{\p}(\varpi_{\sn})}\frac{\rho_{\sn}}{\rho_0}
\eqno{(21)}
$$
Employing the components (1a,b) for $B_\varpi$ and $B_z$ in
$$
\frac{dz}{d\varpi}=\frac{B_z}{B_\varpi}
\eqno{(22)}
$$
and solving the resulting differential equation, subject to the boundary condition $z(\varpi_0)=h_0$, gives the field
line equation
$$
z^2=\varpi^{2/3}_0\left(1+\frac{h^2_0}{\varpi^2_0}\right)\varpi^{4/3}-\varpi^2.
\eqno{(23)}
$$
Using this, together with (1a,b), to evaluate the field strength $B_{\p}=(B^2_\varpi+B^2_z)^{1/2}$ along a poloidal flux
tube yields
$$
B_{\p}=2|B_{z0}|\frac{\varpi^2_0}{\varpi^2}\left[1-\frac{3}{4}\left(\frac{\varpi}{\varpi_0}\right)^{2/3}\right]^{1/2}.
\eqno{(24)}
$$
This shows that $B_{\p}$ increases monotonically as $\varpi$ decreases from the curtain base to the accretor.

 It will be shown that the density decreases from the curtain base to the sonic point, where it has a minimum. It then
increases along the supersonic part of the flow. Hence it follows from (8) that the maximum value of the ratio
$v^2_{\p}/v^2_{\A}$ occurs at the sonic point and that this is related to its value at the curtain base by
$$
\left(\frac{v^2_{\p}}{v^2_{\A}}\right)_{\sn}=\frac{\rho_0}{\rho_{\sn}}\left(\frac{v^2_{\p}}{v^2_{\A}}\right)_0.
\eqno{(25)}
$$
Using (19) and (21) then gives
$$
\left(\frac{v^2_{\p}}{v^2_{\A}}\right)_{\sn}=\left(\frac{\varpi_{\mm}}{\varpi_{\A}}\right)^{7/2}
\left(\frac{\varpi_0}{h_0}\frac{a}{v_{\kp 0}}\right)\frac{h_0}{\delta}\frac{|B_z(\varpi_0)|}{B_{\p}(\varpi_{\sn})}.
\eqno{(26)}
$$
This is significantly less than unity and hence the poloidal speed is well sub-Alfv\'enic everywhere in the curtain flow.
Distortions to the stellar poloidal magnetic field are consequently small and the use of its unperturbed form in the
curtain flow is therefore justified. Equations (7), (8) and (24) show that the field winding ratio $|B_\phi/B_{\p}|$
decreases monotonically from the disruption radius to the stellar surface accretion radius. It is noted that it is the
product $B_\phi B_{\p}$ which is related to angular momentum transfer to the star, and this magnetic stress term increases
with decreasing $\varpi$ through the curtain flow.

\subsection{The toroidal magnetic field and angular velocity}

 In the region just inside the disruption radius, $\varpi_{\mm}$, a transition occurs from the highly diffusive disc
flow to the field controlled curtain flow. In the disc, the winding of the vertical magnetic field by the vertical
shear is balanced by diffusion and the induction equation leads to the surface toroidal field
$$
B_{\phi\s}=|B_z|\frac{\gamma}{\epsilon_{\mm}}\frac{(1-\xi^{3/2}x^{3/2})}{x^{n+1/2}}
\eqno{(27)}
$$
where $x=\varpi/\varpi_{\mm}$, $n>0$ and $\gamma/\epsilon_{\mm}\lesssim1$ corresponds to diffusion processes maintaining
moderate values of field winding (see Campbell 2010). In the curtain flow the creation of $B_\phi$ due to the
variation of $\Omega$ along ${\bf B}_{\p}$ is balanced by its advection due to the poloidal flow, yielding (7) for
$B_\phi$.  

 At $\varpi=\varpi_0=\varpi_{\mm}$ it follows that $x=1$ and (27) gives
$$
\left(\frac{B_{\phi\s}}{|B_z|}\right)_0=\frac{\gamma}{\epsilon_{\mm}}(1-\xi^{3/2}).
\eqno{(28)}
$$
The angular velocity of material must vary from a value of $\Omega_{\kp}(\varpi_{\mm})$ in the disc mid-plane $z=0$
to a value consistent with the curtain flow solution (3) at the base $z=h_0$, with the magnetic field being continuous
acrosss the boundary. Equations (2) and (3) give
$$
\Omega_0=\Omega(\varpi_0,z_0)=\Omega_*-\frac{v_{z0}}{\varpi_{\mm}}\left(\frac{B_{\phi\s}}{|B_z|}\right)_0.
\eqno{(29)}
$$
Noting from (12) that $\Omega_*=\xi^{3/2}\Omega_{\kp\mm}$, and using (28) for $(B_{\phi\s}/|B_z|)_0$, (29) yields
$$
\Omega_0={\bar f}\Omega_*
\eqno{(30)}
$$
where
$$
{\bar f}=1-\frac{\gamma}{\epsilon_{\mm}}\frac{v_{z0}}{v_{\kp\mm}}\frac{(1-\xi^{3/2})}{\xi^{3/2}}.
\eqno{(31)}
$$
Since the last term in (31) is small, ${\bar f}$ is slightly less than unity. It is noted that this vertical transition
in $\Omega$ is necessary since otherwise $B_\phi$ would attain very large negative values at the base of the curtain flow.
Such values would be inconsistent with strong magnetic channelling in the accretion curtain and would be unsustainable, being removed by instabilities and reconnection.

\section{Angular Momentum Transfer}

\subsection{Angular momentum transport in the curtain flow}

 The total rate of transport of angular momentum, per unit poloidal magnetic flux, is given by (5). Since $\beta$ is
conserved along poloidal magnetic field lines, it can be expressed in terms of quantities evaluated at the accretion
curtain base $(\varpi_0, h_0)$. Noting that $B_\phi(\varpi_0,h_0)=B_{\phi\s}(\varpi_0)$, (5) yields
$$
\beta=|\epsilon|\varpi^2_0\Omega_0+\frac{\varpi_0}{\mu_0}B_{\phi\s}(\varpi_0).
\eqno{(32)}
$$
Since $\epsilon=\rho v_z/B_z$, the ratio of the magnetic term to the material term in (32) can be expressed as
$$
\frac{\varpi_0B_{\phi\s}(\varpi_0)}{\mu_0|\epsilon|\varpi^2_0\Omega_0}
=\left(\frac{v^2_{\A}}{v^2_{\p}}\frac{v_z}{v_{\kp}}\frac{B_{\phi\s}}{|B_z|}\right)_0
=\frac{\delta}{\varpi_{\mm}}\left(\frac{B_{\phi\s}}{|B_z|}\right)_0
\left(\frac{\varpi_{\A}}{\varpi_{\mm}}\right)^{7/2} \ll 1,
\eqno{(33)}
$$
where (18) has been used to eliminate $(v^2_{\A}/v^2_{\p})_0$. Hence the last term can be dropped in (32) to give
$$
\beta=\frac{(\rho v_z)_0}{|B_z|_0}\varpi^2_0\Omega_0
=\frac{{\dot M}\varpi^2_0\Omega_0}{4\pi\varpi_{\mm}\delta |B_z|_0},
\eqno{(34)}
$$
where (17) has been employed to eliminate $(\rho v_z)_0$. It follows that $\beta>0$ and this will lead to a positive
accretion torque on the star.

 It is noted that $\Omega$ must change from $\Omega_{\kp\mm}=\Omega_{\kp}(\varpi_{\mm})$ at $z=0$ to $\Omega_0$ 
at the curtain base $z=h(\varpi_0)$. Since $\Omega_0<\Omega_{\kp\mm}$, the transition from the central plane 
of the disc to the accretion curtain base involves a loss of angular momentum per unit mass of 
$\varpi^2_{\mm}(\Omega_{\kp\mm}-\Omega_0)$. The flow in this region changes from horizontal at $z=0$ to vertical at
$z=h(\varpi_0)=h_0$. The angular momentum equation is given by
$$
\frac{\partial}{\partial\varpi}(\varpi\rho v_\varpi\varpi^2\Omega)
+\frac{\partial}{\partial z}(\varpi\rho v_z\varpi^2\Omega)=
\frac{\partial}{\partial\varpi}\left(\rho\nu\varpi^3\frac{\partial\Omega}{\partial\varpi}\right)
+\frac{\partial}{\partial z}\left(\frac{1}{\mu_0}\varpi^2B_\phi B_z\right),
\eqno{(35)}
$$
relating the divergence of the angular momentum flux to the viscous and magnetic torques. Integrating over the region
$0<z<h_0$, noting that $B_\phi$ and $v_z$ vanish at $z=0$, yields
$$
\varpi(\rho v_z)_0\varpi^2\Omega_0-\frac{1}{\mu_0}\varpi^2B_{\phi\s}B_{z\s}=
\frac{d}{d\varpi}\left(\varpi^3\int^{h_0}_0\rho\nu\frac{\partial\Omega}{\partial\varpi}dz
-\varpi^3\int^{h_0}_0\rho v_\varpi\Omega dz\right).
\eqno{(36)}
$$
This transition region of the disc has a radial extent of $\varpi_{\mm}-\delta<\varpi<\varpi_{\mm}$. The disc completely
ends at its inner edge and matter is field dominated here. Hence appropriate inner edge conditions are
$$
\rho(\varpi_{\mm}-\delta,z)=0 \quad {\mbox{and}} \quad 
\left(\frac{\partial\Omega}{\partial\varpi}\right)_{\varpi_{\mm}-\delta}=0.
\eqno{(37{\mbox{a,b}})}
$$
At the outer edge
$$
\left(\frac{\partial\Omega}{\partial\varpi}\right)_{\varpi_{\mm}}=\Omega^\prime_{\kp}(\varpi_{\mm})
=-\frac{3}{2}\frac{\Omega_{\kp\mm}}{\varpi_{\mm}}.
\eqno{(38)}
$$
Integrating (36) over this region, applying (37a,b) and (38), then gives
$$
\left[\varpi_{\mm}(\rho v_z)_0\varpi^2_{\mm}\Omega_0-\frac{1}{\mu_0}\varpi^2_{\mm}B_{\phi\s}B_{z0}\right]\delta
=-\frac{3}{4}\varpi^2_{\mm}\Omega_{\kp\mm}\left(\nu\Sigma\right)_{\varpi_{\mm}}
+\frac{\dot M}{4\pi}\varpi^2_{\mm}\Omega_{\kp\mm},
\eqno{(39)}
$$
where
$$
\Sigma=2\int^{h_0}_0\rho dz.
\eqno{(40)}
$$
and
$$
{\dot M}=-4\pi\int^{h_0}_0\varpi\rho v_\varpi dz
\eqno{(41)}
$$
with ${\dot M}$ being the mass transfer rate into the outer surface at $\varpi=\varpi_{\mm}$. It was shown in Campbell 
(2010) that the boundary condition
$$
\left(\nu\Sigma\right)_{\varpi_{\mm}}=q\frac{\dot M}{3\pi}
\eqno{(42)}
$$
is consistent with disc disruption due to magnetic heating, for $q\lesssim 0.4$. Using this, together with (17) and
(33), in (39) leads to
$$
{\dot M}\varpi^2_{\mm}(\Omega_{\kp\mm}-\Omega_0)=q{\dot M}\varpi^2_{\mm}\Omega_{\kp\mm}
\eqno{(43)}
$$
Noting that $\Omega_0={\bar f}\Omega_*$, with ${\bar f}$ close to unity, and $\Omega_*=\xi^{3/2}\Omega_{\kp\mm}$,
(43) gives the angular momentum balance condition
$$
q=1-\xi^{3/2}.
\eqno{(44)}
$$
This can be satisfied by the values of $\xi$ of most interest, which are nearer to 1 than to 0, since in many observed
systems the star has been spun up to a state in which $\xi$ is well above zero. Small values of $\xi$ may lead to a
break down in steady state conditions.

\subsection{The accretion torque}

 The accretion torque on the star can be evaluated from
$$
T_{\ac}=\int_S\beta{\bf B}_{\p}\cdot d{\bf S},
\eqno{(45)}
$$
where the area $S$ incorporates the regions where the upper and lower curtain flows meet the stellar surface as
ring accretion columns, and the conserved quantity $\beta$ is given by (34). Since the accretion curtains are
symmetric about the mid-plane $z=0$, the total accretion torque will be twice that exerted through the upper
curtain flow. The surface field ratio $B_{\theta\s}/B_{r\s}$, evaluated where the narrow accretion column meets
the star, is given by
$$
\frac{B_{\theta\s}}{B_{r\s}}=\frac{1}{2}\tan\theta_{\ac}
=\frac{1}{2}\tan\left[\sin^{-1}\left(\frac{R}{\varpi_{\mm}}\right)^{1/2}\right],
\eqno{(46)}
$$
where $\theta_{\ac}$ is the angle a tangent to ${\bf B}_{\p\s}$ makes to the vertical and $R$ is the stellar radius. 
For white dwarfs and neutron stars, $R/\varpi_{\mm}\ll 1$ so $\theta_{\ac}$ is small and (46) yields
$B_{r\s}\gg B_{\theta\s}$. Hence the accretion column is very nearly normal to the stellar surface, and a unit normal 
vector to its base is nearly parallel to ${\hat{\bf z}}$. It follows that the magnetic flux through the accretion column base is
$$
\psi_{\mm}=2\pi\varpi_*\delta_*B_{z*},
\eqno{(47)}
$$
where $\delta_*$ is the width of the circular band of cylindrical radius $\varpi_*$ forming the column base. So, allowing for the contribution from the lower accretion column band, the integral in (45) gives
$$
T_{\ac}=2\beta\psi_{\mm}=4\pi\varpi_*\delta_*B_{z*}\beta.
\eqno{(48)}
$$
Then, using conservation of magnetic flux through the accretion curtain to give $\varpi_*\delta_*B_{z*}=
\varpi_{\mm}\delta |B_z|_0$, (48) yields
$$
T_{\ac}=4\pi\varpi_{\mm}\delta|B_z|_0\beta={\dot M}\varpi^2_0\Omega_0,
\eqno{(49)}
$$
where the last equality follows from using (34) for $\beta$. Since ${\bar f}$ is very close to unity, (30) gives
$\Omega_0=\Omega_*$ and hence the accretion torque is
$$
{\bf T}_{\ac}={\dot M}\varpi^2_0\Omega_*\,{\hat{\bf z}},
\eqno{(50)}
$$
To first order in $h(\varpi_0)/\varpi_0$, $\varpi_0$ can be replaced by $\varpi_{\mm}$ so,
since $\Omega_*=\xi^{3/2}\Omega_{\kp}(\varpi_{\mm})$, (50) can be expressed as
$$
{\bf T}_{\ac}=\xi^{3/2}{\dot M}\varpi^2_{\mm}\Omega_{\kp}(\varpi_{\mm})\,{\hat{\bf z}}.
\eqno{(51)}
$$
Noting that $\xi<1$, this expression is a factor $\xi^{3/2}$ smaller than the standard form previously adopted for the
accretion torque. The difference arises from the specific angular momentum $\varpi^2_{\mm}(\Omega_{\kp}-\Omega_*)$
transferred to the disc via viscous stresses as material flows into the accretion curtain. For values of $\xi$ in the
lower part of the range $0.7\lesssim\xi<1$, $T_{\ac}$ can be significantly reduced.

\section{The Slow Magnetosonic Point}

\subsection{The slow point speed}

 The slow magnetosonic point in the curtain flow, and its relation to the sonic point, is now considered. The energy 
integral given by (9) is conserved along field lines on which $z=z(\varpi)$, so substitution of (4) and (6) for $v_{\p}$ 
and $\Omega$ in (9) leads to the relation
$$
H(\varpi,\rho)=E.
\eqno{(52)}
$$
Using $dH=0$ leads to
$$
\frac{d\rho}{d\varpi}=-\frac{\partial H/\partial\varpi}{\partial H/\partial\rho}.
\eqno{(53)}
$$  
The condition $\partial H/\partial\rho=0$ generally yields $v_{\p}$ equal to the slow and fast magnetosonic speeds
(e.g. Campbell 1997, Mestel 1999). In the case considered here the flow is sub-Alfv\'enic, so only the slow speed 
solution $v_{\p}=v_{\ssl}$ will result. At this point $\partial H/\partial\varpi$ must vanish, and $\rho$ has a 
minimum value here.

 The function $H(\varpi,\rho)$ can be expressed as
$$
H=\frac{\epsilon^2B^2_{\p}}{2\rho^2}+\frac{1}{2}\Omega^2_*\varpi^2\left(\frac{\Omega}{\Omega_*}\right)
\left(\frac{\Omega}{\Omega_*}-2\right)-\frac{GM}{[\varpi^2+z(\varpi)^2]^{1/2}}+a^2\ln\rho.
\eqno{(54)}
$$
Since the flow is well sub-Alfv\'enic, distortions of ${\bf B}_{\p}$ from its unperturbed form will be small. 
Equation (6) gives
$$
\frac{\Omega}{\Omega_*}
=\frac{\Omega_*\varpi^2\rho-\mu_0|\epsilon|\beta}{\Omega_*\varpi^2\rho-\mu_0\epsilon^2\Omega_*\varpi^2}.
\eqno{(55)}
$$
Using this in (54) to form $\partial H/\partial\rho=0$, and then (4) to eliminate $B_{\p}$, leads to
$$
v^2_{\ssl}=a^2-\frac{\Omega^2_*}{\varpi^2_{\ssl}}\left(\frac{\mu_0\epsilon^2}{\rho_{\ssl}}\right)^2
\frac{(\varpi^2_{\ssl}-\beta/|\epsilon|\Omega_*)^2}{(1-\mu_0\epsilon^2/\rho_{\ssl})^3}.
\eqno{(56)}
$$
Employing (4), (8), (30) with ${\bar f}=1$, and (34) yields
$$
\frac{v^2_{\ssl}}{a^2}=1-\frac{\Omega^2_*\varpi^2_{\ssl}}{a^2}\left(\frac{v_{\p}}{v_{\A}}\right)^4_{\ssl}
\frac{[(\varpi_0/\varpi_{\ssl})^2-1]^2}{[1-(v_{\p}/v_{\A})^2_{\ssl}]^3}.
\eqno{(57)}
$$
Since $(v_{\p}/v_{\A})^2_{\ssl}\ll 1$, this term can be dropped in the denominator of the last term in (57). It will
be shown that this term is small, so (57) yields the first order expression
$$
v_{\ssl}=a\left(1-\frac{\Omega^2_*\varpi^2_{\ssl}}{2a^2}\left(\frac{v_{\p}}{v_{\A}}\right)^4_{\ssl}
\left[\frac{\varpi^2_0}{\varpi^2_{\ssl}}-1\right]^2\right).
\eqno{(58)}
$$
Hence $v_{\ssl}$ is slightly below the sound speed $a$.

\subsection{The slow point coordinates}

 The condition $\partial H/\partial\varpi=0$ determines the coordinates of the slow magnetosonic point. Equations (4),
(8), (23), (30), (34), (54) and (55) then lead to
$$
v^2_{\ssl}\left(\frac{\varpi}{B_{\p}}\left|\frac{dB_{\p}}{d\varpi}\right|\right)_{\ssl}
+\Omega^2_*\varpi^2_{\ssl}\left[1-2\left(\frac{v^2_{\p}}{v^2_{\A}}\right)_{\ssl}
+\frac{\varpi^4_0}{\varpi^4_{\ssl}}\left(\frac{v^2_{\p}}{v^2_{\A}}\right)^2_{\ssl}\right]
\left[1-\left(\frac{v^2_{\p}}{v^2_{\A}}\right)_{\ssl}\right]^{-2} 
-\frac{2}{3}\left(\frac{\varpi_0}{\varpi_{\ssl}}\right)^{2/3}v^2_{\kp 0}=0.
\eqno{(59)}
$$
Since $(\varpi |dB_{\p}/d\varpi|/B_{\p})_{\ssl}\sim 1$, $(v^2_{\p}/v^2_{\A})_{\ssl}\ll 1$, 
$v^2_{\ssl}/\Omega^2_*\varpi^2_{\ssl}\ll 1$ and $v^2_{\ssl}/v^2_{\kp 0}\ll 1$, (59) reduces to
$$
\Omega^2_*\varpi^2_{\ssl}-\frac{2}{3}\left(\frac{\varpi_0}{\varpi_{\ssl}}\right)^{2/3}v^2_{\kp 0}=0.
\eqno{(60)}
$$
Noting that, to first order in $h_0/\varpi_0$, $\varpi_0=\varpi_{\mm}$ together with $\Omega_*=\xi^{3/2}\Omega_{\kp\mm}$,
(60) leads to
$$
\varpi_{\ssl}=\left(\frac{2}{3}\right)^{3/8}\frac{1}{\xi^{9/8}}\varpi_0.
\eqno{(61)}
$$
and then the field line equation (23) yields
$$
z_{\ssl}=\left(\frac{2}{3}\right)^{1/4}\frac{1}{\xi^{3/4}}\left[1-\left(\frac{2}{3}\right)^{1/4}\frac{1}{\xi^{3/4}}
+\frac{h^2_0}{\varpi^2_0}\right]^{1/2}\varpi_0.
\eqno{(62)}
$$
Equations (61) and (62) give the coordinates of the slow magnetosonic point.

 The condition $\varpi_{\ssl}<\varpi_0$ is necessary to ensure that the slow point lies outside the disc, in the curtain 
flow. Applying this to (61) leads to a critical value for $\xi$ given by
$$
\xi_{\crit}=\left(\frac{2}{3}\right)^{1/3}.
\eqno{(63)}
$$
The slow point will lie above the disc for $\xi>\xi_{\crit}$. The term in $h^2_0/\varpi^2_0$ must be retained in (62) to ensure that $z_{\ssl}\rightarrow h$ as $\xi\rightarrow \xi_{\crit}$.

 The foregoing analysis has shown that for $\xi \leq \xi_{\crit}$ the slow magnetosonic point merges with the disc
surface. The consequences of this result are discussed below.

\subsection{The slow magnetosonic and sonic points}

 The quantity $v^2_{\p}/v^2_{\A}$ gives the ratio of the poloidal kinetic energy density to the poloidal magnetic
energy density. This ratio is significantly less than unity throughout the curtain flow and hence distortions of
the stellar magnetic field are small, so ${\bf B}_{\p}$ is close to its unperturbed form. Since $v^2_{\p}/v^2_{\A}\ll 1$,
(8) shows that the $\rho$-dependent term in the denominator of (7) is negligible and hence
$$
B_\phi=\frac{\mu_0\beta}{\varpi}-\mu_0|\epsilon|\Omega_*\varpi.
\eqno{(64)}
$$
This expression, together with (24) for $B_{\p}$, shows that $B_\phi/B_{\p}$ decreases monotonically from the curtain
flow base to the star. Since $(B_\phi/B_{\p})_0<1$, it follows that field winding is small as expected for a well 
sub-Alfv\'enic flow. The function $\Omega(\varpi,\rho)$ is obtained from (6), (8), (30) and (31) as
$$
\Omega=\Omega_*\left[1-\frac{\gamma}{\epsilon_{\mm}}\frac{v_{z0}}{v_{\kp\mm}}\frac{(1-\xi^{3/2})}{\xi^{3/2}}
\frac{\varpi^2_0\rho_0}{\varpi^2\rho}\right].
\eqno{(65)}
$$
Since $v_{z0}/v_{\kp\mm}\ll 1$, and $\varpi^2_0\rho_0/\varpi^2\rho$ decreases with $\varpi$ inside the sonic point, the
$\rho$-dependent term in (65) is always small. This corresponds to inertia being small relative to magnetic force and so
material is effectively channelled by the magnetic field and nearly corotates with the star, with $\Omega$ approaching
$\Omega_*$ from below as $\varpi^2_0\rho_0/\varpi^2\rho$ decreases with $\varpi$. This corresponds to $\Omega$ being weakly
dependent on $\rho$ so, ignoring its $\rho$ dependence, (54) leads to the critical point condition
$$
\frac{\partial H}{\partial\rho}=-\frac{\epsilon^2B^2_{\p}}{\rho^3}+\frac{a^2}{\rho}
=-\frac{v^2_{\p}}{\rho}+\frac{a^2}{\rho}=0,
\eqno{(66)}
$$
yielding $v_{\p}=a$. This corresponds to the correction term in (58) being ignorable, due to the smallness of
$(v^2_{\p}/v^2_{\A})_{\ssl}$, with the slow magnetosonic and sonic points merging to this degree of accuracy.
 
 With $\Omega$ close to $\Omega_*$, the second term in (54) becomes the centrifugal potential which, together with
the gravitational potential, dominates the poloidal kinetic energy term. The condition $(\partial H/\partial\varpi)_\rho
=0$ then becomes
$$
\frac{d\psi}{d\varpi}=\left(\frac{\partial\psi}{\partial\varpi}\right)_z
+\left(\frac{\partial\psi}{\partial z}\right)_\varpi\frac{dz}{d\varpi}
=\left(\frac{\partial\psi}{\partial\varpi}\right)_z
+\frac{B_z}{B_\varpi}\left(\frac{\partial\psi}{\partial z}\right)_\varpi=0,
\eqno{(67)}
$$
with the effective potential
$$
\psi=-\frac{GM}{(\varpi^2+z^2)^{1/2}}-\frac{1}{2}\Omega^2_*\varpi^2,
\eqno{(68)}
$$
and $z=z(\varpi)$ along the poloidal magnetic field. This is equivalent to the condition
$$
{\bf B}_{\p}\cdot\nabla\psi=0,
\eqno{(69)}
$$
Equation (69) shows that at this critical point the components of the stellar gravitational force and centrifugal force along  ${\bf B}_{\p}$ cancel. For $\varpi<\varpi_{\ssl}$ gravity becomes increasingly dominant along ${\bf B}_{\p}$ 
and the supersonic magnitude of ${\bf v}_{\p}$ approaches free-fall values. The foregoing analysis illustrates that
strong magnetic field channelling of the flow leads to the slow magnetosonic and sonic points merging to high accuracy.
 
 The condition
$$
H(\varpi_0,\rho_0)=H(\varpi_{\sn},\rho_{\sn})
\eqno{(70)}
$$
can be used to determine the ratio $\rho_{\sn}/\rho_0$. Using (54) for $H(\varpi,\rho)$, together with $\varpi_{\sn}=
\varpi_{\ssl}$ and $z_{\sn}=z_{\ssl}$ in (61) and (62), yields
$$
\ln\left(\frac{\rho_{\sn}}{\rho_0}\right)=-\frac{1}{2}\left(1-\frac{v^2_{z0}}{a^2}\right)
-\frac{1}{2}\frac{v^2_{\kp 0}}{a^2}f(\xi),
\eqno{(71)}
$$
where
$$
f(\xi)=\xi^3-4\left(\frac{2}{3}\right)^{3/4}\xi^{3/4}+2.
\eqno{(72)}
$$
The term in $v^2_{z0}/a^2$ makes a small contribution in (71) and can be ignored to a good approximation, giving
$$
\frac{\rho_{\sn}}{\rho_0}=\exp\left[-\frac{v^2_{\kp 0}}{2a^2}f(\xi)\right].
\eqno{(73)}
$$
Using this in (21), and employing (24) for $B_{\p}$, leads to
$$
\frac{v_{z0}}{a}=\frac{1}{2}\left(\frac{2}{3}\right)^{3/4}\frac{1}{\xi^{9/4}}
\left[1-\frac{3}{4}\left(\frac{2}{3}\right)^{1/4}\frac{1}{\xi^{3/4}}\right]^{-1/2}
\exp\left[-\frac{v^2_{\kp 0}}{2a^2}f(\xi)\right].
\eqno{(74)}
$$
For $\xi\leq\xi_{\crit}$, the sonic point merges with the curtain flow base and $v_{z0}=a$.

 This illustrates the relevance of the results of the previous section. For stellar rotation rates having $\xi$ well
above $\xi_{\crit}$, there is a significant effective potential barrier that material must climb in order to flow
through the sonic point and become gravitationally accelerated towards the stellar accretion poles. For values of
$\Omega_*$ that correspond to $\xi\leq\xi_{\crit}$, this barrier is effectively removed due to the sonic point reaching
the disc surface. This occurs because for these rotation rates the lower centrifugal force presents less of a barrier
to the flow. 

\section{Discussion}

\subsection{Summary}

 The structure of an accretion curtain flow, formed as a result of the magnetic disruption of a disc, has been considered.
Disruption occurs due to growing magnetic heating which causes the disc to expand vertically over a narrow radial region
and matter is fed into the accretion curtain. The strong stellar magnetic field channels the flow, which subsequently
passes through a sonic point beyond which it is gravitationally accelerated along the poloidal magnetic field to reach
nearly free-fall speeds before it accretes onto the star's surface. The angular momentum transferred to the star results
in an accretion torque which causes the star to spin-up. This transfer occurs through the curtain flow via associated
magnetic stresses.

 Matching the magnetic field across the boundary connecting the diffusive disc flow and magnetically channelled flow
allows the angular velocity distribution in the curtain flow to be calculated. The resulting sonic point coordinates
are found, with the density reaching a minimum value at this critical point. The sonic point position depends on the
stellar rotation rate, and a critical rate results below which the sonic point is shown to merge with the disc surface.
Above this rotation rate, material leaving the disc must surmount a potential barrier in order to pass through the sonic
point and be gravitationally accelerated towards the star along field lines. Pressure gradient forces allow material to
reach the sonic point, beyond which they make an increasingly small contribution to the acceleration. As the critical
rotation rate is approached very little thermal assistance is required. 

 The accretion torque is calculated due to the transfer of angular momentum to the star via magnetic stresses operating
through the curtain flow. It is shown that the angular momentum balance is maintained by some angular momentum being
transferred back to the disc through the thin transition region over which disruption occurs. The enhanced radial gradient
in the angular velocity occurring in this region facilitates the transfer, via viscous stresses. As a result, the accretion torque can be significantly lower than in the standard model in which all the angular momentum of material is transferred from the inner region of the disc to the star. 

 A dipolar stellar magnetic field is considered here, since this will be the dominant multipole in neutron star and white
dwarf accretors. However, the present work should have relevance to stars in which higher multipole components are
significant, such as in T Tauri systems. Non-axisymmetric effects, such as azimuthal exchange of angular momentum, and
time-dependent effects will tend to average out over a rotation period of the star. The stellar spin evolution will be
a response to such an averaged torque.

\subsection{Comparison with other work}

 The above results can be compared with previous models of angular momentum transfer from the inner part of a magnetically
disrupted disc to the accreting star. The present work agrees with the Ghosh \& Lamb (1978, 1979a,b) picture of a thin
inner radial boundary layer over which disruption occurs. This was shown in Campbell (2010) to be due to rapidly increasing
magnetic dissipation in the disc, which is related to the steeply increasing value of $B^2_{\p}$ inside the corotation radius. The consequent heating of the disc causes it to expand vertically over a narrow region, with the vertical component of stellar gravity and the toroidal field magnetic pressure gradient not being able to balance increases in the vertical thermal pressure gradient. This confirms the idea of Campbell (1992) that disc disruption is a thermally related mechanism involving break down of the disc vertical equilibrium. A transition from the disc flow to the magnetically channelled curtain flow occurs through this inner region.

 The present work also agrees with the Ghosh \& Lamb picture that the angular velocity of disc material changes from a
Keplerian value to the stellar value across the inner boundary layer. Ghosh \& Lamb assumed that all the angular momentum
of material in this layer is transferred to the star via the field-channelled accretion flow, and this leads to the
standard form of ${\dot M}\varpi^2_{\mm}\Omega_{\kp}(\varpi_{\mm})$ for the magnitude of the accretion torque. However, it
is shown here that the increased values of $|d\Omega/d\varpi|$ in the bounday layer allow some angular momentum to be fed
back into the disc. This corresponds to the angular momentum lost by matter as it flows from the central plane of the disc,
where $\Omega=\Omega_{\kp}(\varpi_{\mm})$, to the base of the accretion curtain, where $\Omega$ is close to $\Omega_*$
Consequently, the accretion torque is reduced by a factor of $\xi^{3/2}$, where $\xi=\varpi_{\mm}/\varpi_{\co}$. This result is consistent with the continuity of the magnetic field across the curtain flow base, and it avoids the problem 
of a large value of $|B_\phi|$ just above the disc. Such values of $|B_\phi|$ would be unsustainable and inconsistent with field channelling in the curtain flow. 

 The work of Erkut \& Alpar (2004) indicates that $\Omega$ may approach $\Omega_*$ over a broader region than a boundary
layer for values of $\xi$ in the lower part of the range $0<\xi<1$. However, their analysis did not incorporate the thermal problem and the effects of magnetic heating. As outlined above, magnetic heating effects were shown by Campbell (2010) to lead to rapid disruption of the disc inside $\varpi_{\co}$, and are related to the steep increase in $B^2_{\p}$. 
Nevertheless, this low $\xi$ regime merits further investigation. As noted in Section 4.1, steady state conditions may break down at such low stellar rotation rates.

 The model developed by Shu et al (1994a,b) and Ostriker \& Shu (1995) also has a reduced accretion torque. However, this
is due to angular momentum removal via a magnetically influenced local wind developing just beyond the disruption region,
rather than via viscous transport as in the present model. It is interesting to note that the work of Campbell (2011)
illustrates that as the stellar spin rate approaches the equilibrium value, $P_{\eq}$, at which the accretion and magnetic
disc torques cancel, a narrow region of disc thickening occurs beyond $\varpi_{\co}$. This would aid the production of a
local magnetically channelled wind and hence, for $P$ close to $P_{\eq}$, an accretion curtain and wind flow may co-exist.
Close to the equilibrium state values of the rotation parameter typically lie in the range $0.9 <\xi < 1$ and so 
$\varpi_{\mm}$ lies not far beneath $\varpi_{\co}$. As outlined in Section 1.2, the Shu et al model has disc disruption close to $\varpi_{\co}$, with a narrow X-type region generating a curtain-type inflow and a wind outflow. Hence the 
present model could share some features of the Shu et al model in the limit $P\rightarrow P_{\eq}$.

\section{Acknowledgement}

The author thanks the reviewer for helpful comments which led to improvements in the presentation of the paper.

\end{document}